\RequirePackage[displaymath]{lineno} % Display line numbers
\documentclass[aps,twpcolumn,showpacs,byrevtex,prl,reprint]{revtex4-1}
\usepackage{epsfig}
\usepackage{graphicx}% Include figure files
\usepackage{dcolumn}% Align table columns on decimal point
\usepackage{bm}% bold math
\usepackage{overpic}
\usepackage{subfigure}
\usepackage{float}
\usepackage{color}
\usepackage{amsmath}
\usepackage{mathcomp}
\usepackage{mathrsfs}
\usepackage{multirow}
\usepackage{rotating}
\usepackage{hyperref}
\usepackage{threeparttable}

\begin{document}
\normalsize
\parskip=5pt plus 1pt minus 1pt

%\linenumbers
\title{ \boldmath Observation of an Altered $a_{0}(980)$ Line-shape in $D^{+} \rightarrow \pi^{+}\eta\eta$ due to the Triangle Loop Rescattering Effect}
\vspace{-1cm}

\author{
   \begin{small}
    \begin{center}
    M.~Ablikim$^{1}$, M.~N.~Achasov$^{4,c}$, P.~Adlarson$^{77}$, X.~C.~Ai$^{82}$, R.~Aliberti$^{36}$, A.~Amoroso$^{76A,76C}$, Q.~An$^{73,59,a}$, Y.~Bai$^{58}$, O.~Bakina$^{37}$, Y.~Ban$^{47,h}$
% ...
             \\
         \vspace{0.2cm}
   (BESIII Collaboration)\\
\vspace{0.2cm} {\it
$^{1}$ Institute of High Energy Physics, Beijing 100049, People's Republic of China\\
% ...
$^{4}$ Budker Institute of Nuclear Physics SB RAS (BINP), Novosibirsk 630090, Russia\\
% ...
$^{36}$ Johannes Gutenberg University of Mainz, Johann-Joachim-Becher-Weg 45, D-55099 Mainz, Germany\\
$^{37}$ Joint Institute for Nuclear Research, 141980 Dubna, Moscow region, Russia\\
% ...
$^{47}$ Peking University, Beijing 100871, People's Republic of China\\
% ...
$^{58}$ Southeast University, Nanjing 211100, People's Republic of China\\
$^{59}$ State Key Laboratory of Particle Detection and Electronics, Beijing 100049, Hefei 230026, People's Republic of China\\
% ...
$^{73}$ University of Science and Technology of China, Hefei 230026, People's Republic of China\\
% ...
$^{76}$ University of Turin and INFN, (A)University of Turin, I-10125, Turin, Italy; (B)University of Eastern Piedmont, I-15121, Alessandria, Italy; (C)INFN, I-10125, Turin, Italy\\
$^{77}$ Uppsala University, Box 516, SE-75120 Uppsala, Sweden\\
% ...
$^{82}$ Zhengzhou University, Zhengzhou 450001, People's Republic of China\\
\vspace{0.2cm}
$^{a}$ Deceased\\
% ...
$^{c}$ Also at the Novosibirsk State University, Novosibirsk, 630090, Russia\\
% ...
$^{h}$ Also at State Key Laboratory of Nuclear Physics and Technology, Peking University, Beijing 100871, People's Republic of China\\
% ...
}\end{center}
\vspace{0.4cm}
\end{small}
}

\affiliation{}
\vspace{-4cm}
%\date{\today}

\begin{abstract}
Using 20.3~${\rm fb}^{-1}$ of $e^{+}e^{-}$ collision data taken with the BESIII detector at the center-of-mass energy 3.773~GeV, 
we report the first amplitude analysis of the hadronic decay $D^{+} \rightarrow \pi^{+}\eta\eta$. 
The intermediate process $D^{+} \to a_{0}(980)^{+}\eta, a_{0}(980)^{+} \to \pi^{+}\eta$ is observed and is found to be the only component and its branching fraction is measured to be $(3.67\pm0.12_{\mathrm{stat.}}\pm 0.06_{\mathrm{syst.}})\times 10^{-3}$. 
Unlike the $a_{0}(980)$ line-shape observed in the decays of charmed mesons to $a_{0}(980)\pi$ and in the decay $D^{0} \to a_{0}(980)^{-}e^{+}\nu_{e}$,  where the low-mass side of the $a_0(980)$ is wider than the high-mass side, %due to the couple channel effect, 
the $a_{0}(980)$ line-shape in $D^{+} \to a_{0}(980)^{+}\eta$ is found to be significantly altered, with the high-mass side being wider than the low-mass side. 
We establish that the $a_0(980)$ line-shape arises from the triangle loop rescattering of $D^+ \to \bar{K}_0^*(1430)^0K^+ \to a_0(980)^+ \eta$ and $D^+ \to K_0^*(1430)^+\bar{K}^0 \to a_0(980)^+ \eta$ with a significance of 5.8$\sigma$. This is the first experimental confirmation of the triangle loop rescattering effect. 
\end{abstract}
\clearpage
\newpage
\maketitle
\clearpage
\newpage

%The fundamental theory of strong interactions is Quantum Chromodynamics (QCD). 
%
%At the low-energy region, it exhibits non-perturbative properties. 
Quantum Chromodynamics (QCD) is the fundamental theory of strong interactions, and exhibits non-perturbative properties in the low-energy region.
A significant manifestation of this property is the critical role of hadronic loop contributions in hadronic reactions. 
As a notable example, the invariant mass spectrum of the $\pi^0\pi^0$ system in the decay $K^\pm \to \pi^\pm\pi^0\pi^0$ displays a distinct cusp structure at the $\pi^+\pi^-$ threshold, which can only be explained by a two-point loop diagram with an intermediate $\pi^+\pi^-$ pair~\cite{NA482:2005wht}. 
Moreover, the special singularities associated with triangle loop diagrams, known as triangle singularities, are crucial for interpreting experimental results. 
For instance, in the large isospin-breaking process $\eta(1405/1475) \to \pi\pi\pi$ observed by the BESIII collaboration~\cite{BESIII:2012aa}, a consistent theoretical explanation suggests the involvement of a triangle loop diagram with $K^*\bar{K}K$ on the three legs of the triangle.
In this case, the triangle singularity enhances the isospin-breaking effect \cite{Wu:2011yx,Aceti:2012dj,Wu:2012pg}. 

\begin{figure}[htbp]
\begin{center}
\begin{minipage}[b]{0.30\textwidth}
\epsfig{width=0.98\textwidth,clip=true,file=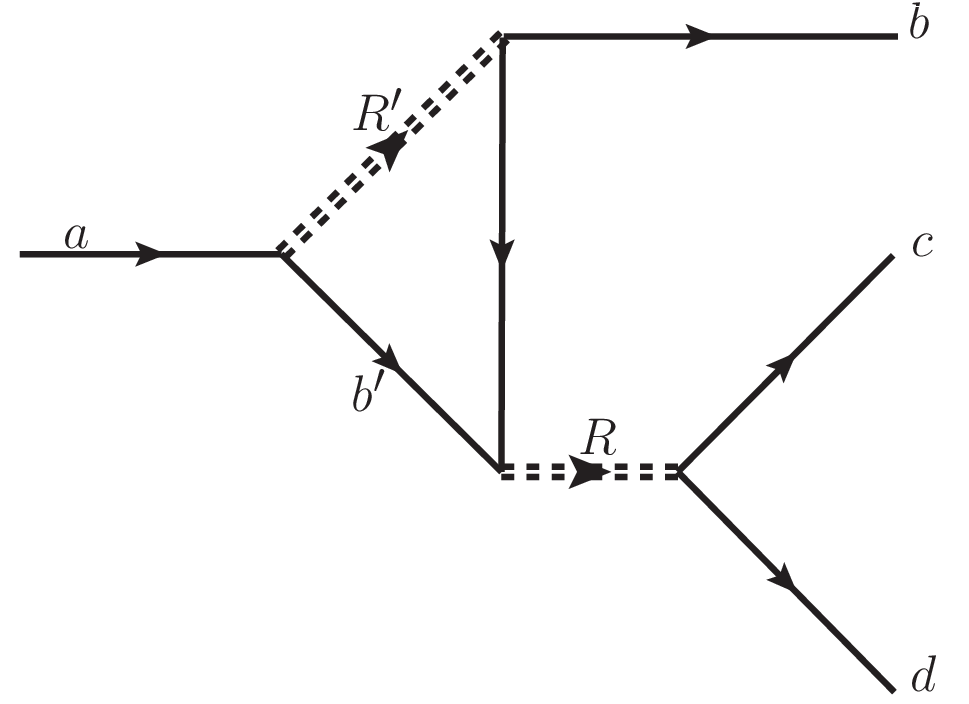}
\end{minipage}
\caption{The triangle loop rescattering diagram for the decay $a \to Rb, R \to cd$.}
\label{fig:triangleloop}
\end{center}
\end{figure}

Apart from these particular thresholds and kinematic effects caused by loop diagrams, 
%Beyond these special threshold and kinematic effects brought by loop diagrams, 
a more intuitive representation of the triangle diagram is as follows. 
As shown in Fig.~\ref{fig:triangleloop}, when the coupling of a parent particle to an intermediate channel $R'b'$ is comparable to that of another channel $Rb$, the contribution of the triangle loop must be considered in experimental data analysis.
Ignoring these contributions could lead to inaccuracies in determining the physical parameters of the resonance state $R$. 
In theoretical analyses, such triangle diagrams are widely discussed, especially in the context of the hadronic molecular candidates~\cite{Guo:2017jvc}. 
Nevertheless the contributions of such loop diagrams have predominantly been discussed in theory, and only to a limited extent in the amplitude analyses applied directly to experimental data.

Recently, the COMPASS collaboration incorporated triangle diagrams into their experimental analysis, and found that a triangle singularity is likely responsible for the signal orignally interpreted as the the $a_1(1420)$ particle~\cite{COMPASS:2020yhb}.
Similarly, the LHCb collaboration has considered the impact of triangle diagrams in the analysis of the $\Lambda_b^0 \to J/\psi p K^-$ decay~\cite{LHCb:2019kea}.
However, these analyses only suggest that triangle loop diagrams might play a significant role in reactions, without conclusively demonstrating their contribution.
Notably, there remains a lack of experimental evidence directly demonstrating the crucial effects of triangle loop diagrams on the extraction of resonance states.
%
%Hence finding a way to clearly identify triangle loop contributions and establishing a partial wave analysis method that incorporates loop diagram calculations is urgent and important.
%Hence clearly experimental identifications of triangle loop and establishing their contributions is urgent and critical. 
Therefore, it is essential to develop a clear method for identifying triangle loop contributions and to establish an amplitude analysis framework that incorporates loop diagram calculations.

This Letter aims to clarify the contributions of triangle loop diagrams in the hadronic decay $D^+ \to \pi^+ \eta\eta$.
Our analysis reveals that contributions from $\bar{K}_0^*(1430)^0K^+ \to a_0(980)^+\eta$ and $K_0^*(1430)^+\bar{K}^0 \to a_0(980)^+\eta$ rescattering, where the $\bar{K}_0^*(1430)^0K^+\bar{K}^0$ and $K_0^*(1430)^+\bar{K}^0K^+$ form the three legs of a triangle loop diagram, respectively, significantly alter the line-shape of the $a_0(980)$.
We study the mass spectrum of the $a_{0}(980)$ via an amplitude analysis of $D^{+} \to \pi^{+}\eta\eta$ by using 20.3~${\rm fb}^{-1}$ of $e^{+}e^{-}$ collision data~\cite{Ablikim:2013ntc,BESIII:2015equ,BESIII:2024} recorded with the BESIII detector at the center-of-mass energy 3.773 GeV. Charge conjugation is implied throughout this Letter.
This Letter represents the first empirical validation of the triangle loop diagrams in experimental analyses, laying a foundation for future research in this area.

The BESIII detector and the upgraded multi-gap resistive plate chambers used in the time-of-flight systems are described in Refs.~\cite{detector,MRPC} in detail, respectively.
Since the $D^{+}$ and $D^{-}$ are produced together in pairs, the double tag (DT) method~\cite{MARK-III:1985hbd} is employed to suppress backgrounds. Six tag channels 
$D^{-} \to K^{+}\pi^{-}\pi^{-} (\pi^{0})$, $D^{-} \to K^{0}_{S}\pi^{-} (\pi^{0})$, $D^{-} \to K^{0}_{S}\pi^{-}\pi^{-}\pi^{+}$, and $D^{-} \to K^{+}K^{-}\pi^{-}$ are used.
Signal MC samples with $\psi(3770) \to D^{+}D^{-}$, $D^{-} \to \mathrm{tags}$ and $D^{+} \to \pi^{+}\eta\eta$ are generated, 
where the amplitude model from the fit to the data is used for the signal decay. 
The tracking, particle identification (PID), $K_{S}^{0}$, $\pi^{0}$, and $\eta$ reconstruction are almost identical to those in Ref.~\cite{BESIII:2023exq},
except for the signal window of invariant mass for $\eta$ candidates, which is set to $0.45<M(\gamma\gamma)_{\eta}<0.65$~GeV$/c^{2}$ to improve the $\eta$ reconstruction efficiency. 
Two variables, $M_{\mathrm{BC}} = \sqrt{E_{\mathrm{beam}}^{2} - |\Vec{P}_{D^{\pm}}|^{2}}$ and 
$\Delta E = E_{D^{\pm}} - E_{\mathrm{beam}}$, are used to identify $D^{\pm}$ mesons,
where $(E_{D^{\pm}},\Vec{P}_{D^{\pm}})$ is the four-momentum of the $D^{\pm}$ meson and $E_{\mathrm{beam}}$ is the beam energy.
For both the tag and signal sides, any candidate with $M_{\mathrm{BC}}<1.83$~GeV$/c^{2}$ or $|\Delta E|>0.1$~GeV is first rejected. The candidate for each tag mode with $\Delta E$ closest to 0 is 
selected, while the signal candidate with 
$M_{\mathrm{BC}}$ closest to the $D$ nominal mass~\cite{Workman:2022ynf} is chosen. 
Backgrounds are studied with an inclusive Monte Carlo (MC) sample simulated with {\sc geant4}~\cite{sim},
which includes all known open-charm decays, charmless decays and initial-state radiative decays to the $J/\psi$ or $\psi(3686)$. All particle decays with known branching fractions are modeled with {\sc evtgen}~\cite{EvtGen}. The remaining unknown charmonium decays are generated with {\sc lundcharm}~\cite{Chen:2000tv}. 

For the tag channels, the $M_{\mathrm{BC}}$ signal windows are defined as $\pm 6$~MeV$/c^{2}$ around the nominal $D^{-}$ mass~\cite{Workman:2022ynf}; 
and the $\Delta E$ windows are set to 3.5 times the corresponding resolutions~\cite{BESIII:2023exq}. 
On the signal side, the $M_{\mathrm{BC}}$ signal window is required to be in the range $[1.860,~1.880]$~GeV$/c^{2}$.
Since the dominant background is from processes with no $\eta$ in the final state, for events used in the amplitude analysis 
a multi-variant analysis~\cite{Hocker:2007ht} is employed in which a Gradient Boosted Decision Tree (BDTG) classifier is developed based on the inclusive MC sample. The BDTG takes five discriminating variables: the natural logarithm of the $\chi^{2}$ of constraining the two-photon pairs to the $\eta$ mass; 
the invariant masses and the cosine of the helicity angles of the photon decaying from $\eta_{1,2}$ with higher energy, 
where the sub-index $1$ and $2$ for $\eta$ represents a more or less energetic $\eta$.
%$M(\gamma\gamma)_{\eta_{1,2}}$, the natural logarithm of the $\chi^{2}$ of constraining the two-photon pairs to the $\eta$ mass ($\chi^{2}(\eta)$) and the cosine of the helicity angle of the photon decaying from $\eta_{1,2}$ with higher energy, where the sub-index $1$ and $2$ for $\eta$ represents a more or less energetic $\eta$.  
Studies of the MC samples show that a requirement on the output of the BDTG retains 78.3\% signals and rejects 89.7\% background.
A further requirement of $|\Delta E|<0.040$~GeV is then imposed.
Furthermore, to ensure all candidate events fall into the physical region of phase space, a kinematic fit is performed, where aside from four-momentum conservation, 
the invariant masses of $M(\gamma\gamma)_{\eta}$ and $M(\pi^{+}\eta \eta)_{D}$ are constrained to individual values quoted from the Particle Data Group (PDG)~\cite{Workman:2022ynf}.
Finally, a sample of 1624 candidates is retained with a purity of $(85.1 \pm 0.9)\%$.

The amplitude analysis is performed using accepted events in data based on an unbinned maximum likelihood fit. The logarithm of the likelihood is constructed as 
\begin{eqnarray}
\begin{aligned}
\ln L = \ln(f_{\mathrm{s}}\Tilde{S}(p) + (1-f_{\mathrm{s}})\Tilde{B}(p)),
\end{aligned}
\end{eqnarray}
where $f_{\mathrm{s}}$ is the signal purity, $p$ is the four-momenta of the final particles, and $\Tilde{S}(p)$ and $\Tilde{B}(p)$ are the probability density functions (PDFs) of signal and background, respectively:
\begin{eqnarray}
\begin{aligned}
\Tilde{S}(p)& = \frac{\epsilon(p)|\mathcal{M}(p)|^{2}R_{3}(p)}{\int{\epsilon(p)|\mathcal{M}(p)|^{2}R_{3}(p)\mathrm{d}p}},\\
\Tilde{B}(p)& = \frac{\epsilon(p)B_{\epsilon}(p)R_{3}(p)}{\int{\epsilon(p)B_{\epsilon}(p)R_{3}(p)\mathrm{d}p}}.	
\end{aligned}
\end{eqnarray}
Here, $R_{3}(p)$ is the three-body phase space factor, $\epsilon(p)$ is the efficiency function, and $B_{\epsilon}(p) = B(p)/\epsilon(p)$.
The $B(p)$ is the function extracted from the simulated background sample. 
The $\mathcal{M}(p)$ is the total amplitude of the signal, which is modeled as the coherent sum of the amplitudes of all intermediate processes, $\mathcal{M}(p) = \sum_{\alpha}{c_{\alpha}e^{i\phi_{\alpha}}A_{\alpha}}$, where $c_{\alpha}$ and $\phi_{\alpha}$ are the magnitude and phase of the $\alpha^{\mathrm{th}}$ amplitude, respectively. 
The amplitude $A_{\alpha}$ is defined as  $A_{\alpha} = P_{\alpha}S_{\alpha}F^{r}_{\alpha}F^{D}_{\alpha}$, where $P_{\alpha}$ is the propagator for an intermediate resonance; $S_{\alpha}$ is the spin factor constructed with the covariant tensor formalism~\cite{Zou:2002ar}; and $F^{r}_{\alpha}$ ($F^{D}_{\alpha}$) is the barrier factor for the intermediate state ($D$ meson). 
For the $S$-wave, both $S_{\alpha}$ and $F_{\alpha}$ factors are 1.0.  
The propagator $P_{\alpha}$ is taken to be a relativistic Breit-Wigner formula for all resonances except the $a_{0}(980)$.  For the $a_0(980)$, a three-channel coupled Flatt\'e formula is used, given by $P_{a_{0}(980)} = 1/[(M_{0}^{2} - s) - i(g_{\eta\pi}^{2}\rho_{\eta\pi} + g_{K\bar{K}}^{2}\rho_{K\bar{K}} + g^{2}_{\pi\eta^{\prime}}\rho_{\pi\eta^{\prime}})]$, where $g_{\pi\eta}$, $g_{K\bar{K}}$ and $g_{\pi\eta^{\prime}}$ are the coupling constants reported in Ref.~\cite{BESIII:2016tqo},  
and $\rho_{\pi\eta}$, $\rho_{K\bar{K}}$ and $\rho_{\pi\eta^{\prime}}$ are phase space factors, calculated as $q/\sqrt{s}$.  
Here, $q$ is the magnitude of the momentum for daughter particles in the rest frame of the $a_{0}(980)$ candidate; 
$M_{0}$ is the mass of the $a_{0}(980)$; and $s$ is the invariant mass squared of the $a_{0}(980)$ candidate.

\begin{figure}[htbp]
\begin{center}
\begin{minipage}[b]{0.22\textwidth}
\epsfig{width=0.98\textwidth,clip=true,file=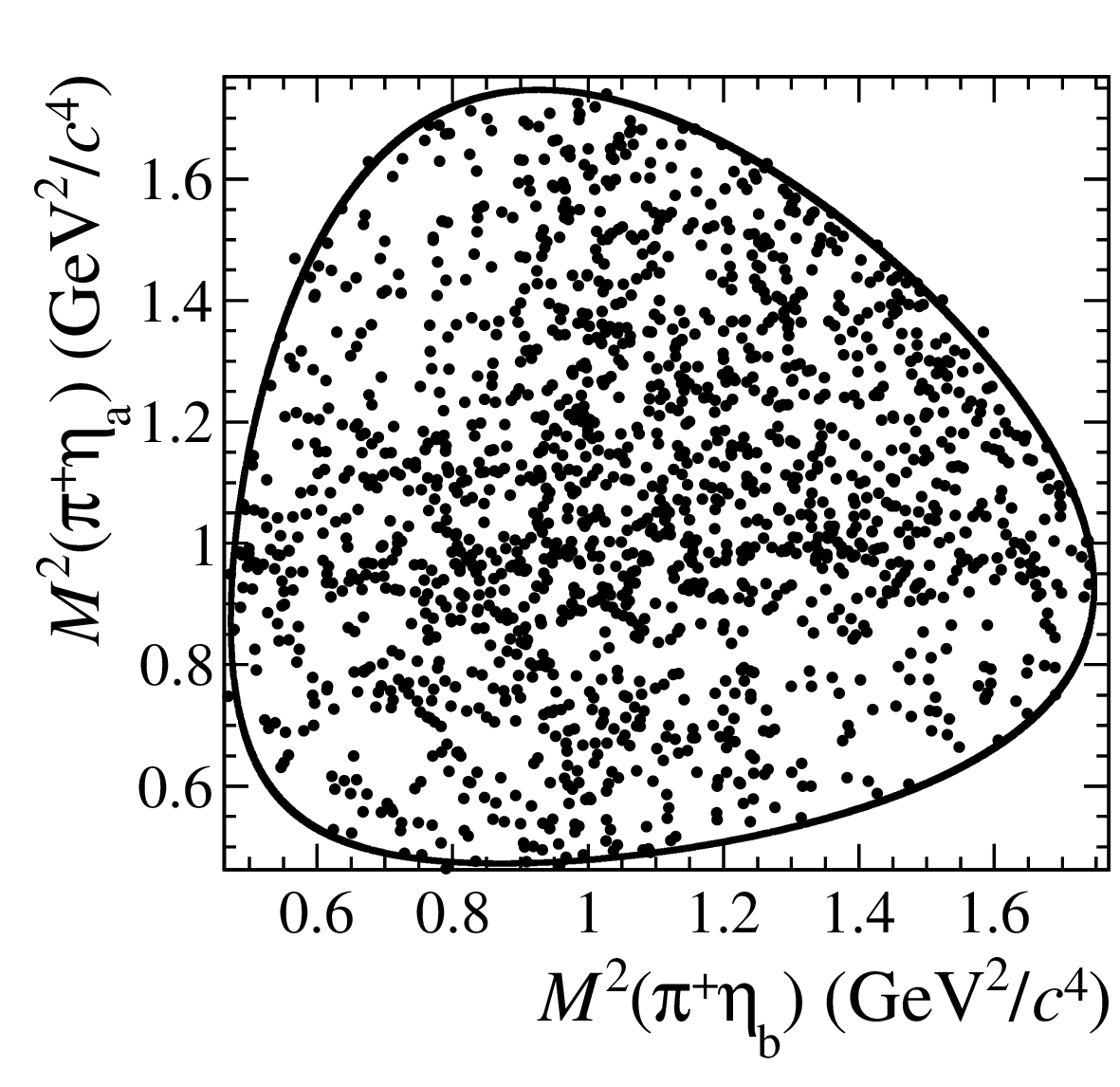}
\put(-15,85){(a)}
\end{minipage}
\begin{minipage}[b]{0.22\textwidth}
\epsfig{width=0.98\textwidth,clip=true,file=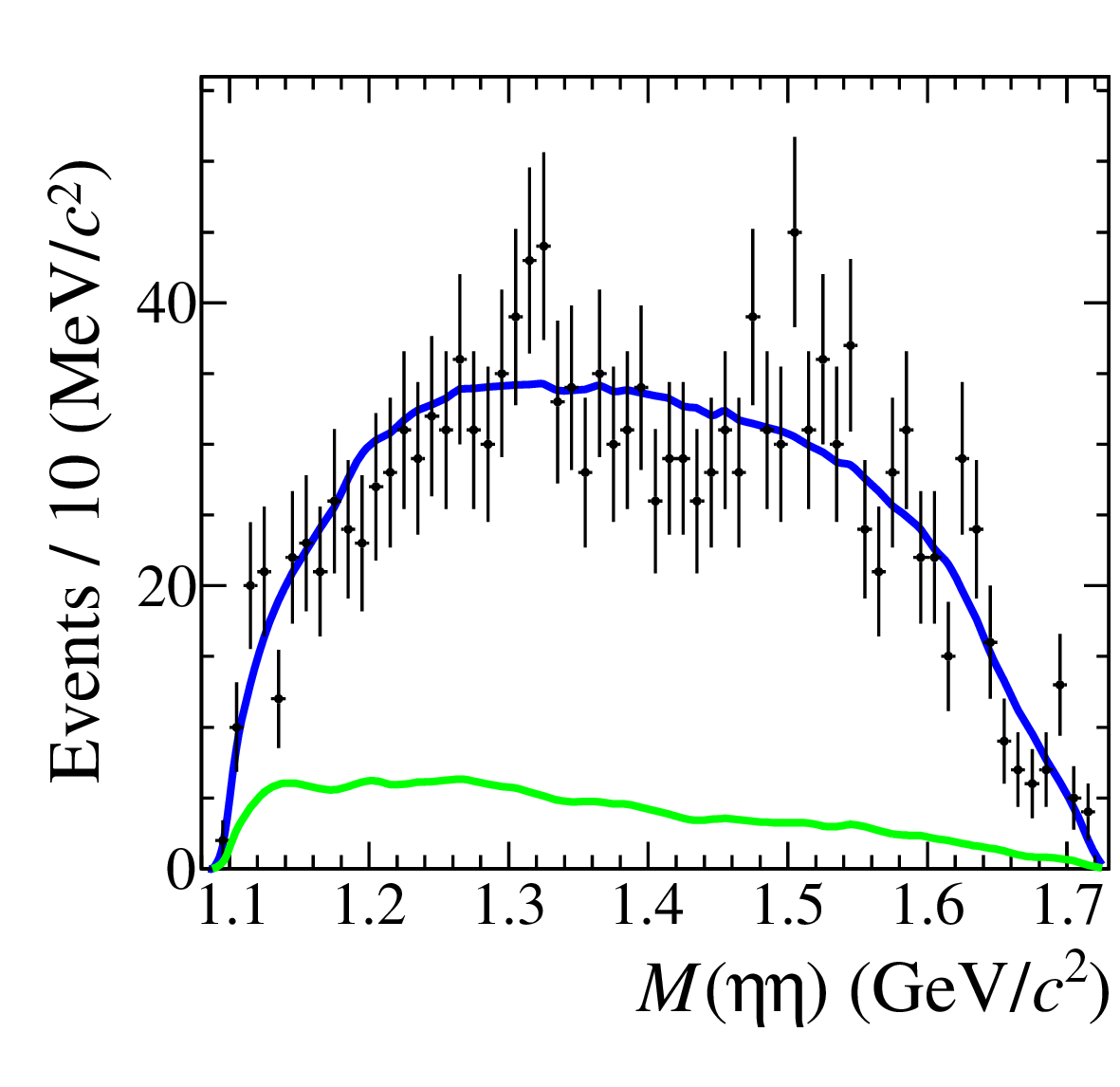}
\put(-15,85){(b)}
\end{minipage}
\begin{minipage}[b]{0.35\textwidth}
\epsfig{width=0.98\textwidth,clip=true,file=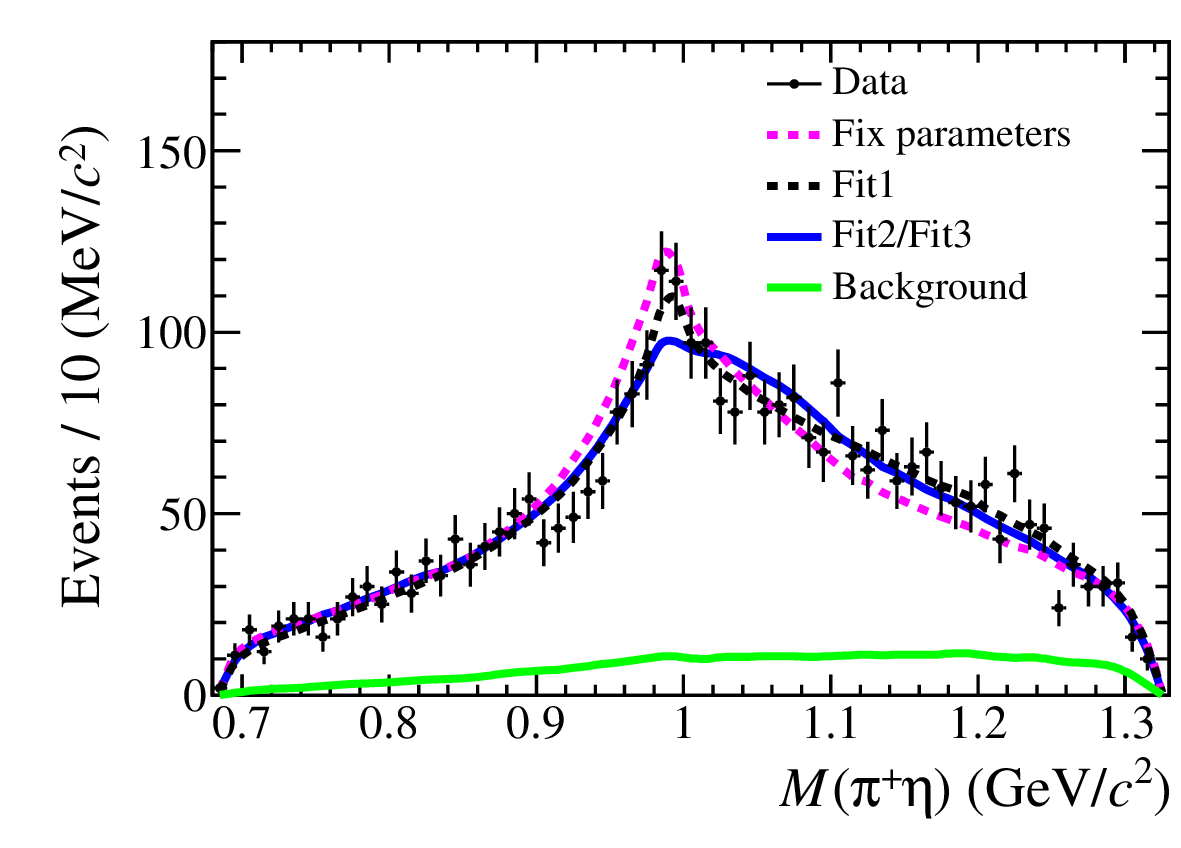}
\put(-125,100){(c)}
\end{minipage}
\caption{The (a) Dalitz plot of the data and the projections on (b) $M(\eta\eta)$ and (c) $M(\pi^{+}\eta)$ are shown. 
In figures (b,c), the dots with errors are data and the green histograms are the background; The blue lines are the fit for Fit2/Fit3 (the lines for Fit2 are almost the same as Fit3). 
In figure (c), the pink dashed line is the projection obtained by fixing the $a_{0}(980)^{+}$ parameters to the previous measurement~\cite{BESIII:2016tqo}, where a global shift around the $a_{0}(980)$ peak is observed; the the black dashed line is the fit for Fit1. 
}
\label{fig:projection}
\end{center}
\end{figure}

For the $M(\eta\eta)$ and $M(\pi^+\eta)$ projections shown in Fig.~\ref{fig:projection} (c,d), only the $a_{0}(980)^{+}$ resonance is observed.   
Therefore, in an initial fit the magnitude and phase of the $D^{+} \to a_{0}(980)^{+} \eta$, $a_{0}(980)^{+} \to \pi^{+}\eta$ amplitude are fixed to be 1.0 and 0.0, respectively.  We then search for additional amplitudes with significance greater than $5\sigma$, where
the significance is calculated using the changes of $\ln L$ and the number of degrees of freedom (NDOF)
when the fit is performed with and without the related amplitude. 
A constant term, representing a uniform distribution in phase space, is the only other amplitude besides the $a_{0}(980)^{+}$ found to have a significance greater than $5\sigma$~\cite{significanceforothers}.
But the model that includes the $a_{0}(980)^{+}$ and the constant term leads to more than 30\% constructive interference.  
Since this is unphysical, we conclude the process $D^{+} \to a_{0}(980)^{+} \eta$, $a_{0}(980)^{+} \to \pi^{+}\eta$ is the only allowed amplitude.

With the coupling constants in the Flatt\'e formula fixed to those reported in Ref.~\cite{BESIII:2016tqo}, the $a_{0}(980)^{+}$ line-shape cannot be well described. 
The fit quality is determined by calculating the $\chi^{2}$ of the fit using an adaptive binning of the $M^{2}(\pi^{+}\eta_{a})$ versus $M^{2}(\pi^{-}\eta_{b})$ Dalitz plot, where $a$ and $b$ denote the two $\eta$ mesons being randomly exchanged. Each bin is required to contain at least 10 events.  This results in $\chi^{2}/\mathrm{NDOF} = 139.0/92$.
For the $a_{0}(980)$ line-shape observed in charmed meson decays to $a_{0}(980)\pi$~\cite{BESIII:2019jjr,BESIII:2024tpv} and the semi-leptonic decay $D^{0} \to a_{0}(980)^{-}e^{+}\nu_{e}$~\cite{BESIII:2024zvp}, 
%for a normal $a_{0}(980)$ line-shape without any other effect, 
the coupled channel effect described by the Flatt\'e formula leads to an asymmetry of the $a_{0}(980)$ line-shape, where its low-mass side is wider than the high-mass side, 
which is different from the Dalitz plot of data shown in Fig.~\ref{fig:projection} (a). 

In an attempt to improve the fit to the $a_{0}(980)$ line-shape, we first allow the $M_{0}$, $g_{\eta\pi}^{2}$ and $g_{K\bar{K}}^{2}$ parameters to float in the fit. 
%the data statistics for $g_{\pi\eta^{\prime}}$ determination is too low. 
The $g_{\pi\eta^{\prime}}$ parameter is always fixed to its value from the previous measurement due to low statistics in the data.
This fit, named ``Fit1'', describes the $a_{0}(980)$ line-shape well, as shown in Fig.~\ref{fig:projection} (c). 
The $M_{0}$, $g_{\eta\pi}^{2}$ and $g_{K\bar{K}}^{2}$ parameters are measured to be 
$(1.074\pm0.020_{\mathrm{stat.}}\pm0.013_{\mathrm{syst.}})~\mathrm{GeV}/c^{2}$, 
$(0.538\pm0.053_{\mathrm{stat.}}\pm0.051_{\mathrm{syst.}})~\mathrm{GeV^{2}}/c^{4}$,
and 
$(0.675\pm0.169_{\mathrm{stat.}}\pm0.163_{\mathrm{syst.}})~\mathrm{GeV^{2}}/c^{4}$, respectively. 
The resulting pole position of the $a_{0}(980)$ is
$[(1.098\pm0.014)-i(0.034\pm0.022)]~\mathrm{GeV}/c^{2}$.  Note that the real part is higher than the $K\bar{K}$ threshold by $(107 \pm 14)~\mathrm{MeV}/c^{2}$, which is inconsistent with previous measurements.
In addition, an input-output (IO) check for fit Fit1 shows there is a possible fit bias in the $a_0(980)$ mass and coupling constants~\cite{IOcorr}.  The IO check is
performed using two sets of data-sized MC with 300 individual samples for each set. 
Here, one set of MC samples is signal only and the other one mixes the background with the same rate as in data. 

%Compared with the PDG result~\cite{Workman:2022ynf}, there is more than 5.5$\sigma$ deviations, thereby ruling out the F1 as a physical result.

Thus, we finally incorporate the triangle loop rescattering diagram shown in Fig.~\ref{fig:triangleloop}. 
According to the PDG, such rescattering can occur through the process $D^{+} \to \bar{K}_{0}^{*}(1430)^{0}K^{+} \to a_{0}(980)^{+}\eta$, and the amplitude can be written as  
$A_{\alpha} = (1 + CA_{\mathrm{loop}})P_{\alpha}$. Here, $C = |C|e^{i\phi_{C}}$ is a complex coefficient, but should be treated as a coupling constant according to the loop amplitude formalism, whose details can be found in the Supplemental Material~\cite{Supplement}. 
Therefore, $\phi_{C}$ is expected to be 0. 
%The formalism of loop amplitude $A_{\mathrm{loop}}$ is given in the supplement.
Allowing both $|C|$ and $\phi_{C}$ to float, in a second fit named ``Fit2'', yields 
$|C| = 0.112 \pm 0.015_{\mathrm{stat.}} \pm 0.048_{\mathrm{syst.}}$, 
and $\phi_{C} = (-0.056\pm0.144_{\mathrm{stat.}}\pm0.085_{\mathrm{syst.}})~\mathrm{rad}$, 
which is consistent with the expectation of $\phi_{C} = 0$. 
Therefore, a third fit named ``Fit3'', with $\phi_{C}$ fixed to 0 is performed, which gives $|C| = 0.113 \pm 0.015_{\mathrm{stat.}} \pm 0.048_{\mathrm{syst.}}$. 
Both Fit2 and Fit3 provide good descriptions of the altered $a_{0}(980)$ line-shape. 
%The values of $|C|$ in F2 and F3 are almost the same. 
The IO check of Fit2 shows about 0.3 standard deviation from the input value 
due to the correlation between $|C|$ and $\phi_{C}$, while for Fit3, the IO check shows good consistency. %Here $\sigma_{\mathrm{stat.}}$ is the statistical uncertainty. 

The significance of the triangle loop re-scattering contribution is determined to be 5.8$\sigma$ by considering the change in $\ln L$ and the number of degrees of freedom
when the fit is performed with and without the loop.  The coupling constants in the Flatt\'e formula are fixed to those reported in Ref.~\cite{BESIII:2016tqo}, and the significance is the lowest one among the fits for Fit2 and Fit3, including the variations performed in the systematic uncertainty studies.  
The Dalitz plots and the projections are shown in Fig.~\ref{fig:projection}. For the two identical $\eta$, the two $M(\pi^{+}\eta)$ spectra are added. 
The $\chi^{2}/\mathrm{NDOF} = 89.3/92$. 

The systematic uncertainties on the floating parameters from the Fit1, Fit2, and Fit3 are determined from various sources as follows.
%For F1, F2 and F3, the systematic uncertainty includes the sources: 
(I)~ For uncertainties from the $a_{0}(980)$ line-shape, we drop the coupled channel $a_{0}(980) \to \pi\eta^{\prime}$ and repeat the fit. (II)~ For uncertainties from the input values of the mass and coupling constants used in the Fit2/Fit3 fits, we shift their values within their uncertainties given in Ref.~\cite{BESIII:2016tqo}. (III)~ For the background ratio, the $f_{s}$ factor is changed within its uncertainty. (IV)~ For the background shape, the largest background component is $D^{+} \to \pi^{+}\pi^{0}\pi^{0}$, and we shifted its ratio within its uncertainty. (V)~ For the fit bias, the uncertainties are obtained from the IO check. 
The detailed results are summarized in Table~\ref{tab:ampsys}.
\begin{table}[htbp]
\begin{center}
\caption{The systematic uncertainties for each floated parameter in Fit1, Fit2 and Fit3. }
\begin{tabular}{c|c|ccccc|c} \hline
\hline
 Fit                &  Parameter        &   I   &  II     &   III  &  IV    &   V    & Total  \\ \hline
\multirow{3}{*}{Fit1} &  $m_{0}~[\mathrm{GeV}/c^{2}]$          & 0.002 &$\cdots$ & 0.003  & 0.000  & 0.012  & 0.013  \\
                    &  $g_{\pi\eta}~[\mathrm{GeV}/c^{2}]^{2}$    & 0.035 &$\cdots$ & 0.006  & 0.000  & 0.036  & 0.051  \\
                    &  $g_{K\bar{K}}~[\mathrm{GeV}/c^{2}]^{2}$   & 0.095 &$\cdots$ & 0.014  & 0.000  & 0.132  & 0.163  \\ \hline
\multirow{2}{*}{Fit2} &  $|C|$            & 0.044 & 0.019 & 0.002  & 0.000  & 0.005  & 0.048  \\
                    &  $\phi_{C}~[\mathrm{rad}]$       & 0.011 & 0.068 & 0.012  & 0.001  & 0.048  & 0.085  \\ \hline
 Fit3                 &  $|C|$            & 0.044 & 0.019 & 0.001  & 0.000  & 0.000  & 0.048  \\ \hline    
 \hline
\end{tabular}
\label{tab:ampsys}
\end{center}
\end{table}

To illustrate the consistency between the loop contribution and the value of the product branching fraction
$\mathcal{B}(D^{+} \to \bar{K}_{0}^{*}(1430)^{0}K^{+})\mathcal{B}(\bar{K}_{0}^{*}(1430)^{0} \to \bar{K}^{0} \eta)$ from the PDG, 
the product $\mathcal{B}(D^{+} \to a_{0}(980)^{+}\eta)\mathcal{B}(a_{0}(980)^{+} \to \pi^{+}\eta)$ is measured. 
Since only one intermediate state contributes to the decay $D^{+} \to \pi^{+}\eta\eta$, we can write $\mathcal{B}(D^{+} \to a_{0}(980)^{+}\eta)$ $\mathcal{B}(a_{0}(980)^{+} \to \pi^{+}\eta)$ $ = \mathcal{B}(D^{+} \to \pi^{+}\eta\eta)$. 

\begin{figure}[htbp]
\begin{center}
\begin{minipage}[b]{0.35\textwidth}
\epsfig{width=0.98\textwidth,clip=true,file=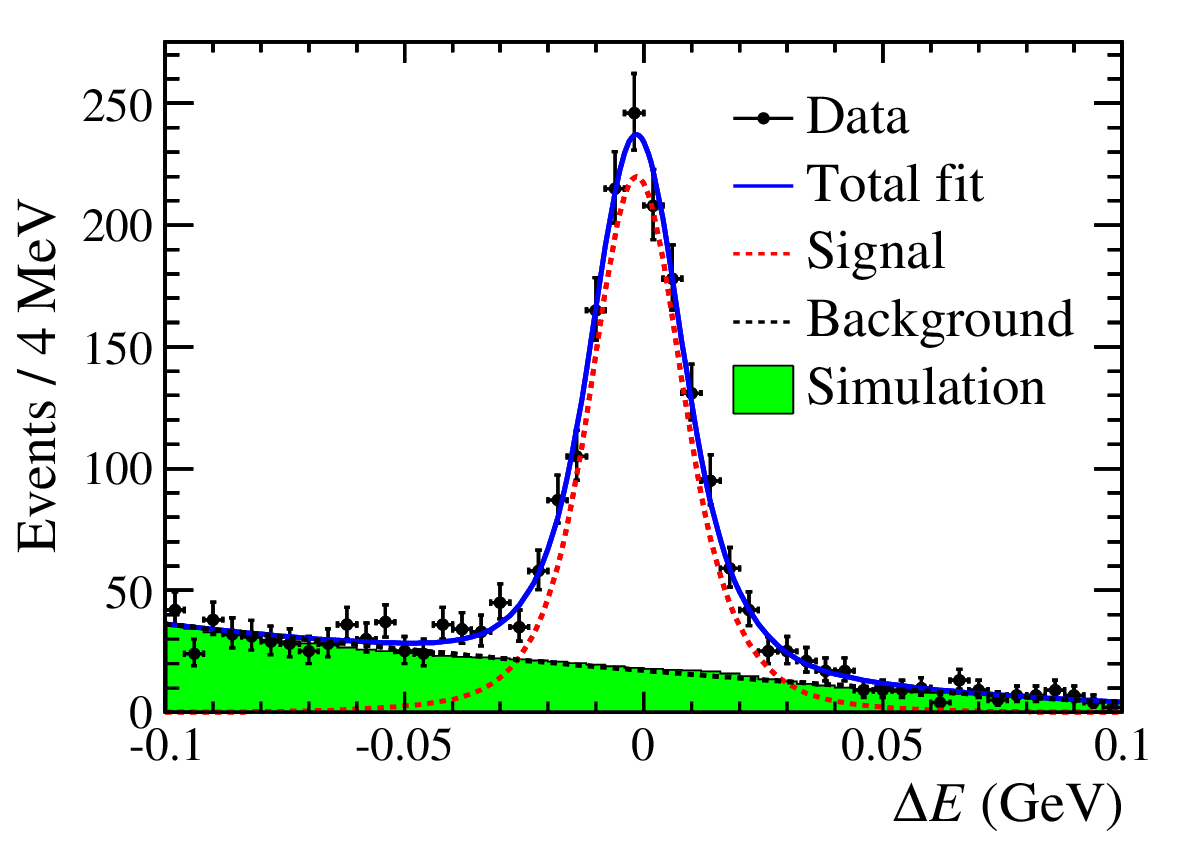}
%\put(-85,80){(b)}
\end{minipage}
\caption{The fit to the $\Delta E$ distribution is shown: 
the dots with errors are the data;
the solid, red dashed and black dashed lines are the total fit, signal and background, respectively;
the green histogram is the background estimated from the inclusive MC sample.}
\label{fig:DT}
\end{center}
\end{figure}

To measure this branching fraction $\mathcal{B}$, we apply the tighter requirements $0.505<M(\gamma\gamma)_{\eta}<0.570$~GeV$/c^{2}$ and $\chi^{2}(\eta)<50$, 
in spite of using the MVA selection. 
The total decay $\mathcal{B}$ is measured by employing the DT method~\cite{MARK-III:1985hbd}, which gives $\mathcal{B} = Y_{\mathrm{DT}}/Y_{\mathrm{ST}}\epsilon_{\mathrm{sig}}\mathcal{B}^{2}_{\mathrm{sub}}$. 
Here, $Y_{\mathrm{ST}}$ is the total single tag (ST) yield, which is $10646901\pm3792$; $\mathcal{B}_{\mathrm{sub}}$ is the $\mathcal{B}(\eta \to \gamma\gamma)$;
in the weighted signal efficiency $\epsilon_{\mathrm{sig}} = \sum_{i}{\frac{Y^{(i)}_{\mathrm{ST}}}{\epsilon^{(i)}_{\mathrm{ST}}}\epsilon^{(i)}_{\mathrm{DT}}}/Y_{\mathrm{ST}}$, 
the $Y^{(i)}_{\mathrm{ST}}$ and $\epsilon^{(i)}_{\mathrm{ST~(DT)}}$ are the ST yield and the ST (DT) efficiencies for the $i^{\mathrm{th}}$ tag channels, respectively.
%which is $\mathcal{B} = Y_{\mathrm{DT}}/(\sum_{i}{\frac{Y^{(i)}_{\mathrm{ST}}}{\epsilon^{(i)}_{\mathrm{ST}}}\epsilon^{(i)}_{\mathrm{DT}}})$. Here $Y_{\mathrm{ST}}$ is the single tag (ST) yields, which is totally $4176863\pm2794$; the $\epsilon_{\mathrm{ST}}$ are the ST efficiencies determined by inclusive MC sample. 
For the DT yield $Y_{\mathrm{DT}}$, they are extracted by fitting the $\Delta E$ distribution without the signal window applied, as shown in Fig.~\ref{fig:DT}. Here the signal function is parameterized with the sum of a bifurcated Gaussian function and a double Gaussian function, 
where the bifurcated Gaussian function is the altered Gaussian function where the left and right widths 
are allowed to be different and the two functions share the same mean value. All the parameters except for the mean value are determined by fitting the signal MC sample. The background function is a $2^{\mathrm{nd}}$-order Chebychev polynomial, which is validated by the inclusive MC sample.
The fit gives $Y_{\mathrm{DT}} = 1506\pm49$.

Using the DT efficiency $\epsilon^{(i)}_{\mathrm{DT}}$, estimated with the signal MC sample, $\mathcal{B}$ is measured to be
$(3.67\pm0.12_{\mathrm{stat.}}\pm 0.06_{\mathrm{syst.}}) \times 10^{-3}$.  
The systematic uncertainties, which total 1.6\% of the mean value, are from the following sources.  
The uncertainties from the PID (0.5\%), tracking (0.5\%), and $\eta$ reconstruction (0.6\%) efficiencies are determined from hadronic DT $D\bar{D}$ events~\cite{BESIII:2024njj};  
the signal shape (0.4\%) and background shape (0.7\%) uncertainties are estimated by the shift of the $Y_{\mathrm{DT}}$ obtained by altering the parameters in the signal shape within uncertainties,   
and fixing the parameters in the polynomial to the fit to the simulated backgrounds derived from the inclusive MC sample, respectively; 
the $M_{\mathrm{BC}}$ window uncertainty is found to be negligible; the ST yield uncertainty (0.3\%) is quoted from Ref.~\cite{BESIII:2024njj}; the fitter performance (0.3\%) is estimated from the inclusive MC sample by employing the same performance as data; the MC generator uncertainty is estimated by varying the parameters in the generator model according to the error matrix obtained from the fit to data and found to be negligible; and the uncertainty from $\mathcal{B}(\eta \to \gamma \gamma)$ (0.9\%) is quoted from the
PDG~\cite{Workman:2022ynf}.

From our measured values of $\mathcal{B}$ and $|C|$, we require $\mathcal{B}(D^{+} \to \bar{K}_{0}^{*}(1430)^{0}K^{+})\mathcal{B}( \bar{K}_{0}^{*}(1430)^{0} \to \bar{K}^{0} \eta)$ to be on the level of $10^{-5}$ for consistency with the triangle rescattering picture. This is on the same order as the PDG value~\cite{Workman:2022ynf}. 

In summary, from the first amplitude analysis of the hadronic decay $D^{+} \to \pi^{+} \eta \eta$, the decay $D^{+} \to a_{0}(980)^{+} \eta$ is observed for the first time. 
In this process, the $a_{0}(980)$ line-shape is found to deviate from the shape expected from the Flatt\'e formalism. 
%After excluding all other possible solutions, 
The fit model including a constant term introduces large constructive interference, and the fit model with floating Flatt\'e parameters results in the pole 
of the $a_{0}(980)$ being far away from the $K\bar{K}$ threshold. 
We conclude that this discrepancy comes from the triangle loop rescattering effect.
This is the first experimental observation of a triangle loop rescattering effect, which has been widely discussed by theorists in the past decades. 
This measurement directly confirms this critical effect in line-shape investigations. 
In the future, experimental analyses will need to carefully consider such kinds of effects, especially in the construction of fit models to extract the parameters of resonant states. 

The BESIII Collaboration thanks the staff of BEPCII (https://cstr.cn/31109.02.BEPC) and the IHEP computing center for their strong support. This work is supported in part by National Key R\&D Program of China under Contracts Nos. 2023YFA1606000, 2020YFA0406400, 2020YFA0406300, 2023YFA1606704; National Natural Science Foundation of China (NSFC) under Contracts Nos. 12205384, 12175239, 12221005, 11635010, 11735014, 11935015, 11935016, 11935018, 12025502, 12035009, 12035013, 12061131003, 12192260, 12192261, 12192262, 12192263, 12192264, 12192265, 12221005, 12225509, 12235017, 12361141819; the Chinese Academy of Sciences (CAS) Large-Scale Scientific Facility Program; the CAS Center for Excellence in Particle Physics (CCEPP); Joint Large-Scale Scientific Facility Funds of the NSFC and CAS under Contract No. U1832207; CAS under Contract No. YSBR-101; 100 Talents Program of CAS; the Excellent Youth Foundation of Henan Scientific Commitee under Contract No.~242300421044; The Institute of Nuclear and Particle Physics (INPAC) and Shanghai Key Laboratory for Particle Physics and Cosmology; Agencia Nacional de Investigación y Desarrollo de Chile (ANID), Chile under Contract No. ANID PIA/APOYO AFB230003; German Research Foundation DFG under Contract No. FOR5327; Istituto Nazionale di Fisica Nucleare, Italy; Knut and Alice Wallenberg Foundation under Contracts Nos. 2021.0174, 2021.0299; Ministry of Development of Turkey under Contract No. DPT2006K-120470; National Research Foundation of Korea under Contract No. NRF-2022R1A2C1092335; National Science and Technology fund of Mongolia; National Science Research and Innovation Fund (NSRF) via the Program Management Unit for Human Resources \& Institutional Development, Research and Innovation of Thailand under Contract No. B50G670107; Polish National Science Centre under Contract No. 2019/35/O/ST2/02907; Swedish Research Council under Contract No. 2019.04595; The Swedish Foundation for International Cooperation in Research and Higher Education under Contract No. CH2018-7756; U. S. Department of Energy under Contract No. DE-FG02-05ER41374.
%\clearpage


\begin{thebibliography}{99}
\bibitem{significanceforothers} The significance for the other possible amplitudes are all less than 3$\sigma$.
\bibitem{IOcorr} In the result for Fit1, the corrections from IO are addressed.
\end{thebibliography}
\end{document}